\documentclass[openany,preprint,superscriptaddress,amsmath,amssymb]{revtex4-2}

\usepackage{graphicx}

\usepackage{color}

\usepackage[dvipsnames]{xcolor}

\usepackage{upgreek}

\usepackage{float}

\usepackage{upgreek}

\usepackage{pdfpages}

\usepackage{placeins}

\newcommand{\degC}[0]{\ensuremath{^{\circ}}C}

\newcommand{\degree}{\ensuremath{^{\circ}}}

\makeatletter
\AtBeginDocument{\let\LS@rot\@undefined}
\makeatother

\begin{document}

\title{Methods for color center preserving hydrogen-termination of diamond}

\author{D. J. McCloskey}

\affiliation{School of Physics, University of Melbourne, Parkville, VIC 3010, Australia}

\author{D. Roberts}

\affiliation{School of Science, RMIT University, Melbourne, VIC 3001, Australia}

\author{L.V.H. Rodgers}
\affiliation{Department of Electrical and Computer Engineering, Princeton University, Princeton, New Jersey 08544, USA}

\author{Y. Barsukov}

\affiliation{Princeton Plasma Physics Laboratory, Princeton University, Princeton, New Jersey 08540, USA}

\author{I. D. Kaganovich}

\affiliation{Princeton Plasma Physics Laboratory, Princeton University, Princeton, New Jersey 08540, USA}

\author{D. A. Simpson}

\affiliation{School of Physics, University of Melbourne, Parkville, VIC 3010, Australia}

\author{N. P. de Leon}

\affiliation{Department of Electrical and Computer Engineering, Princeton University, Princeton, New Jersey 08544, USA}

\author{A. Stacey}

\affiliation{School of Science, RMIT University, Melbourne, VIC 3001, Australia}

\affiliation{Princeton Plasma Physics Laboratory, 100 Stellarator Road, Princeton, NJ 08540, USA}

\author{N. Dontschuk}

\affiliation{School of Physics, University of Melbourne, Parkville, VIC 3010, Australia}
 
\begin{abstract}
\noindent Chemical functionalization of diamond surfaces by hydrogen is an important method for controlling the charge state of near-surface fluorescent color centers, an essential process in fabricating devices such as diamond field-effect transistors and chemical sensors, and a required first step for realizing families of more complex terminations through subsequent chemical processing. In all these cases, termination is typically achieved using hydrogen plasma sources which can etch or damage the diamond as well as deposited materials or embedded colour centers. This work explores alternative methods for lower-damage hydrogenation of diamond surfaces, specifically the annealing of diamond samples in high-purity, non-explosive mixtures of nitrogen and hydrogen gas, and the exposure of samples to microwave hydrogen plasmas in the absence of intentional stage heating. The effectiveness of these methods are characterized by x-ray photoelectron spectroscopy, and comparison of the results to density-functional modelling of the surface hydrogenation energetics implicates surface oxygen ligands as the primary factor limiting the termination quality of annealed samples. Finally, photoluminescence spectroscopy is used to verify that both the annealing and reduced sample temperature plasma methods are non-destructive to near-surface ensembles of nitrogen-vacancy centers, in stark contrast to plasma treatments which use heated sample stages.
\noindent 

\end{abstract}
\maketitle

\section*{Introduction}

\noindent Hydrogen-terminated diamond surfaces exhibit a number of intriguing electronic properties such as a negative electron affinity\cite{ristein_surface_2006}, conductivity via the formation of a surface-confined two-dimensional hole gas (2DHG) \cite{maier_origin_2000}, and sensitivity of the hole density/surface potential to solvated ion concentrations and/or pH\cite{rezek_intrinsic_2007,ristein_hydrogen-terminated_2008}. High quality hydrogenated surfaces can provide a direct coupling between charge at the diamond surface and the electrostatic environment within the shallow bulk material below, a mechanism which has been exploited to control the charge state and fluorescence properties of near-surface colour centers such as nitrogen-vacancy (NV) and, more recently, silicon-vacancy (SiV) defects \cite{zhang_neutral_2023,hertkorn_vacancy_2019,grotz_charge_2012}. This control has been exploited to detect redox activity via NV fluorescence for DNA sensing \cite{krecmarova_label-free_2021}, to enable voltage imaging using dense NV ensembles for use in probing electrophysiological and electrochemical processes \cite{mccloskey_diamond_2022}, and to stabilize individual neutrally-charged SiV centers \cite{zhang_neutral_2023} which could be exploited for sensing \cite{zhang_optically_2020} or long-range photonic quantum communications \cite{huang_hybrid_2021}. Hydrogenated diamond surfaces are also highly homogeneous, making them an ideal starting point for the surface functionalization processes required to introduce tethering ligands for label-free optical biosensors \cite{kayci_multiplexed_2021, raymakers_diamond_2019} and nanoscale magnetic resonance investigation of organic and biomolecular targets using near-surface NV centers \cite{rodgers_diamond_2024,janitz_diamond_2022}. Finally, incorporation of NV centers near active electronic devices allows for direct measurement of internal electric fields and current flow \cite{broadway_spatial_2018}. Situating NV centers in close proximity to the 2DHG may therefore assist in future designs of high-power, high-frequency electronics based on hydrogenated diamond \cite{yang_vector_2020, oing_monolithic_2021, sasama_high-mobility_2021}.\\
\newline
\noindent The standard method employed to form high quality hydrogen terminations is to heat the diamond substrate above 600\,\degC\ while exposing it to hydrogen plasma \cite{dankerl_diamond_2009}. However, these processes have been found to lead to irreversible passivation of nitrogen-vacancy fluorescence even tens of micrometers below the surface \cite{stacey_depletion_2012,findler_indirect_2020}. This loss of fluorescence, particularly limiting for experiments requiring the formation or repeated restoration of high quality hydrogen terminations of near-surface NV containing diamond, motivates the pursuit of alternative hydrogenation methods which can avoid this deleterious effect. Most commonly employed techniques are largely unsatisfactory for this purpose: Electrochemical hydrogen termination requires conductive boron doped diamond which substantially reduces the population of the negatively-charged NV centers required for spin-based sensing protocols \cite{yap_properties_2018,groot-berning_passive_2014,freitas_luminescence_1994}, while annealing in pure hydrogen atmospheres \cite{seshan_hydrogen_2013,manfredotti_about_2010,michaelson_dissociative_2014} requires specialised equipment to mitigate explosion hazards.\\
\newline
\noindent Here, we explore two alternative methods for color-center preserving hydrogenation of diamond: (1) Forming gas (FG) annealing, which has been recently demonstrated to preserve coherence properties of NV centers within 10\,nm of the diamond surface \cite{rodgers_diamond_2024} and enable reversible charge state control of shallow SiV centers \cite{zhang_neutral_2023}, in which the sample is annealed in a non-explosive mixture of hydrogen and nitrogen gas ($<$ 17\,\% H\textsubscript{2} by volume) at atmospheric pressure, and (2) termination by exposure to hydrogen plasma without intentional sample stage heating, which we refer to here as cold plasma termination. X-ray photoelectron spectroscopy (XPS) measurements of surface oxygen content were used to evaluate the quality of the resulting terminations. These revealed low levels of oxygen on cold plasma treated samples and a monotonic temperature dependence of the termination quality (oxygen content) on FG annealed samples. Optimal FG termination quality was observed at annealing temperatures above 900\,\degC. \textit{In situ} XPS measurements of the surface oxygen content for a sample annealed in ultra-high vacuum (UHV) also indicate that thermal removal of oxygen species saturates at around 950\,\degC, in support of prior work suggesting that hydrogen atmosphere does not chemically attack the the oxygenated surface directly \cite{seshan_hydrogen_2013}. We used density-functional and transition-state modelling to explore the termination process and found that the reconstructed diamond (100) surface should undergo near-complete hydrogenation at any temperature above 600\degC\ on exposure to hydrogen atmosphere, a much lower temperature than that required for the FG annealing process. We conclude therefore that abstraction of terminating oxygen species, not thermal or catalytic cracking of atmospheric molecular hydrogen, is the primary mechanism limiting the termination quality of FG annealed samples. Finally, photoluminescence (PL) spectroscopy was used to confirm the suitability of both FG annealing and cold plasma exposure for preserving ensembles of shallow NV centers, while heated sample plasma exposure was found to permanently decrease NV PL emissions in line with expectations.
\section{Methods}
\subsection{Sample Preparation}
\noindent Double-side polished (100) oriented optical-grade single-crystal diamond wafers (EDP) were used for forming gas annealing and plasma termination experiments. The sample used for NV PL measurements was a (100) oriented double-side polished electronic-grade single-crystal diamond wafer (Element Six) implanted with \textsuperscript{14}N\textsuperscript{+} ions at an energy of 3\,keV with a 7\degree incidence angle and a dose of 10\textsuperscript{13}cm\textsuperscript{-2} (Coherent Ion Implantation). Following implantation, the sample was annealed in FG at 950\degC\ for 4 hours to create a near-surface layer of NV centers with a conversion efficiency of nitrogen to NV in the range of 1\%-5\% \cite{pezzagna_creation_2010,tetienne_spin_2018} at an expected mean depth of 5.25\,nm \cite{broadway_spatial_2018}. The sample was then cleaned by acid boiling (1:20 w/w mixture of sodium nitrate and sulphuric acid) for 30 minutes followed by a 10 minute immersion in hot piranha solution (1:4 v/v mixture of hydrogen peroxide and sulphric acid) and ultrasonic cleaning in acetone, isopropyl alcohol, and deionised water. Prior to hydrogen-termination, all samples were oxygen-terminated by exposure to ozone (Samco UV-1). This same ozone treatment procedure was used for each subsequent oxygen termination of the NV containing sample in order to reliably assess damage to the NV layer.
\subsection{Plasma Hydrogen Termination}
\noindent For hydrogen-termination in plasma, a microwave-plasma enhanced chemical vapor deposition reactor (ASTeX) with a molybdenum stage was used to expose the sample to a pure hydrogen plasma using similar recipes to those reported previously for controlling the charge state of near-surface NV ensembles \cite{hauf_chemical_2011,dankerl_diamond_2009}. For heated sample plasma hydrogenation, this consisted of a pressure of 10\,torr, H\textsubscript{2} flow rate of 500\,sccm, a stage temperature of 800\,\degC, and a microwave power of 400\,W for 10 minutes, after which the stage was allowed to cool to room temperature in flowing hydrogen. For unheated sample  "cold plasma hydrogenation", a pressure of 10\,torr, H\textsubscript{2} flow rate of 500\,sccm, and microwave power of 400\,W was employed for either 5, 10, or 50 minutes. A short 3-second plasma strike period using 1200\,W RF power was necessary in these cases. A maximum stage temperature of 120\,\degC\ was recorded during cold plasma termination processes.
\subsection{Thermal Hydrogen Termination}
\begin{figure*}[h]
\centering
\includegraphics[width=\textwidth]{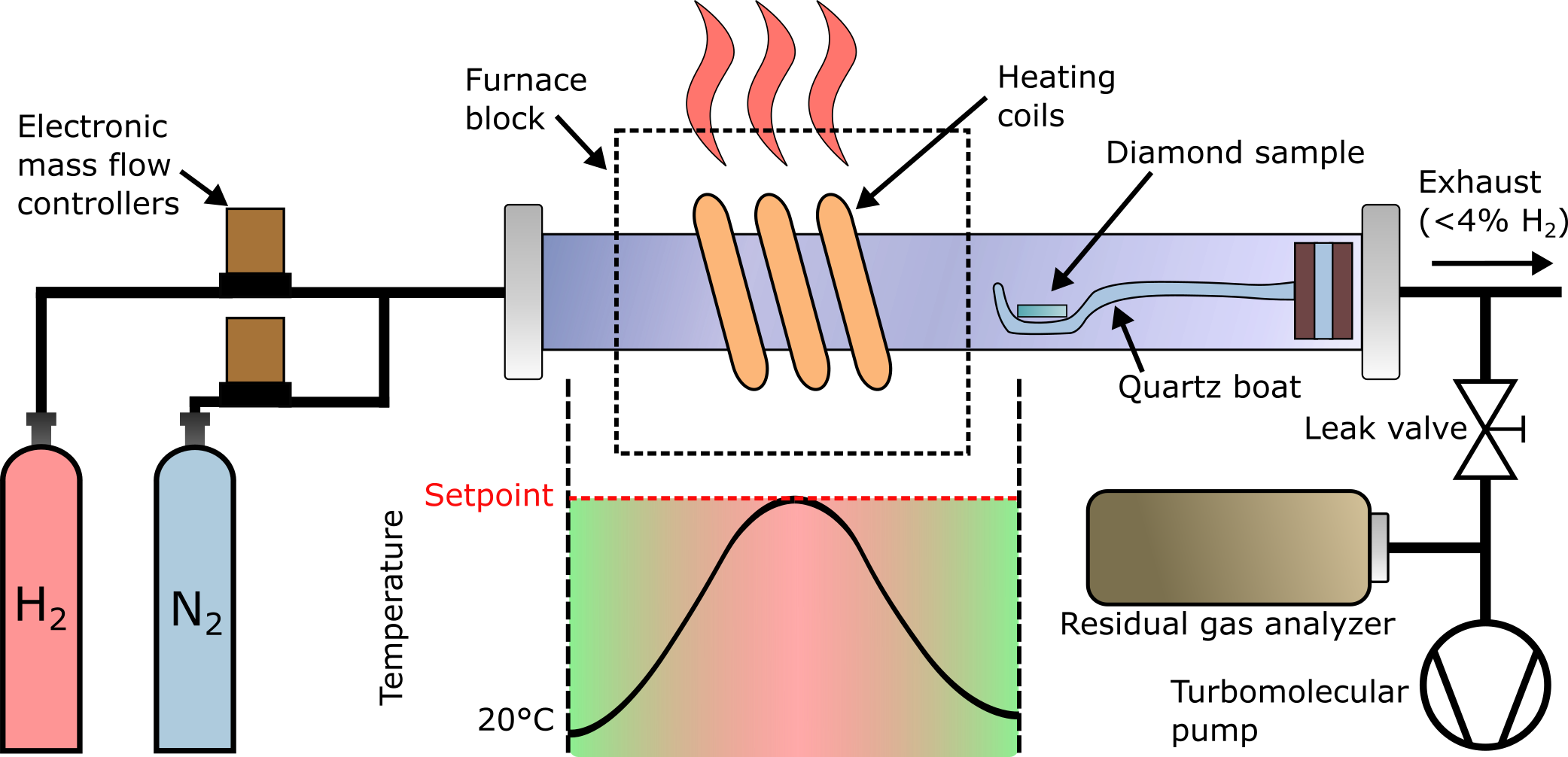}
\caption{Diagram of tube furnace apparatus used for forming gas annealing experiments.}
\label{furnacediagram}
\end{figure*}

\noindent Forming gas annealing was performed in the quartz tube furnace apparatus illustrated in Figure \ref{furnacediagram}. Diamond samples were placed in a quartz boat which could be inserted and removed from the furnace block by a magnetic coupling. We used ultra high purity ($>$99.999\%) hydrogen and nitrogen gas sources slightly over atmospheric pressure to limit the oxygen attack of samples at elevated temperatures. A flow of 7\,sccm of H\textsubscript{2} and 150\,sccm of N\textsubscript{2} was utilized, with gas exhausted from the tube through a paraffin bubbler. The sample was not inserted into the hot zone of the furnace until the partial pressures of oxygen gas and water vapour were measured to be less than 10\,ppm and 100\,ppm respectively using a residual gas analyzer (Stanford Research Systems) connected to the exhaust line through a leak valve. Purging required around 90 minutes for the oxygen and water vapour levels to reduce to the specified concentrations. After this, the sample was moved into the hot zone and left for one hour before being rapidly withdrawn to the room temperature segment of the tube. Samples were allowed to cool to room temperature ($\approx15$~min) under flowing gas before removal.
\subsection{X-ray Photoelectron Spectroscopy}
\noindent XPS measurements following hydrogenation were performed using an AXIS Supra spectrometer (Kratos). Measurements were performed on `as-terminated' samples and following sample cleaning using an \textit{in situ} argon cluster ion source (Kratos Minibeam 6) to remove atmospheric surface adsorbates without heating and verify the absence of other surface contaminants. The argon cluster ion source utilized an accelerating voltage of 10\,keV, an average argon cluster size of 3000 atoms, and an ion charge of +1. An areal dosage of 5.2$\pm$0.4 ions/nm\textsuperscript{2} was used for cleaning the vast majority of samples with the exception of those treated by cold hydrogen plasma, which were cleaned with a dose of 3.9$\pm$0.3 ions/nm\textsuperscript{2}. Termination quality was assessed by comparing the ratio of integrated areas of the fitted O1s and C1s peaks.
\textit{In situ} XPS measurements monitoring thermal oxygen removal were performed in a custom-built UHV chamber using an ASPECT spectrometer (Sigma) while the sample was subjected to a linear heating ramp from room temperature to approximately 920\,\degC\ at a rate of 1\,\degC/minute using a ceramic heating element. Here, the removal of surface oxygen was assessed by monitoring the background-subtracted O1s peak area over the duration of the experiment.
\subsection{Photoluminescence Spectroscopy}
\noindent PL measurements were made with a Renishaw inVia Raman/PL microscope in an upright configuration using a 50$\times$ objective lens (0.5\,NA). Excitation was provided by 532\,nm continuous-wave laser spot focused to the sample surface and delivering a power of 2.68\,mW. Collected emissions were passed through a 550\,nm long-pass filter and 1200 line/mm reflection grating before projection onto a silicon CCD detector (Renishaw) which was thermo-electrically cooled to -70\,\degC. Spectra were acquired using 3 accumulations of 10 second exposures. Intensity was normalised between runs to measurements of the silicon Raman scattering line at 520\,cm\textsuperscript{-1} from a silicon reference sample with fixed crystallographic orientation relative to the illumination polarization. Diamond samples were visually aligned to the same crystallographic orientation prior to measurement.
\subsection{Surface Reaction Modeling}
\noindent Using the Gaussian 16 software package \cite{frisch2016gaussian}, we performed density functional theory (DFT) calculations to estimate the activation energy barriers and temperature dependence of 10 possible reactions (Supporting Information) which describe hydrogenation of the reconstructed diamond (100) surface by both atomic and molecular hydrogen, the thermal desorption of terminating hydrogen atoms, and the abstraction of terminating hydrogen by hydrogen radicals. The activation energies were used in conjunction with transition state theory and the Eyring equation \cite{eyring_activated_1935} to compute (also using the Gaussian 16 package) the temperature-dependent rate constants for all 10 reactions, from which the steady-state fractions of three possible diamond surface configurations (hydrogen dimer, hydrogen monomer, and reconstructed/bare diamond surface) were calculated as a function of temperature and the rate of ambient molecular hydrogen dissociation.
\section{Results}
\subsection{Termination Quality}
\noindent Two identical sets of 12 optical-grade diamond wafers were prepared using the procedures described above. The samples from each set underwent forming gas annealing at 12 different furnace temperatures over a range from 600\,\degC\ to 1100\,\degC. Pairs of samples from the two sets were annealed simultaneously in order to capture sample-to-sample variation under identical processing conditions. After annealing, hydrogenation of diamond surfaces was confirmed by measurement of surface conductivity using a four-point probe. Conductivity commensurate with high nitrogen content diamond was observed on all samples besides those annealed below 700\,\degC. We also measured six argon-cleaned reference samples, two oxygen-terminated with ozone treatment, two hydrogen-terminated using hydrogen plasma treatment with a heated stage, and two hydrogen-terminated using the cold hydrogen plasma treatment recipe for 10 minutes (Methods).\\
\newline
\noindent The surface oxygen coverage of the FG annealed samples determined by XPS in Figure \ref{XPSdata} a) exhibits a decreasing trend with increased annealing temperature, which is consistent with increasing surface coverage of hydrogen. The trend saturates at temperatures above 850\,\degC, resulting a final oxygen coverage greater than that obtained by standard plasma exposure, but less than that obtained via cold plasma termination. Argon cleaning reduced the detected oxygen component across all samples, as would be expected for removal of atmospheric adsorbates without damage to the terminating species. Comparing to the heated plasma-hydrogenated control samples, it is evident that achieving similar termination quality via FG annealing is possible at temperatures above 900\,\degC, provided the observed run-to-run variability can be reduced.\\
\newline
\noindent Next, we turn our attention to understanding the mechanism by which molecular hydrogen terminates the diamond surface. One ambiguity in the above experiment is whether hydrogen plays a key chemical role in removing oxygen groups from the diamond surface, or whether it simply serves to passivate dangling surface bonds. To investigate this, we performed \textit{in situ} XPS measurements of a diamond sample subjected to a linear temperature ramp in a UHV environment, plotted in Figure \ref{XPSdata} b). It is evident from this experiment that heating to 950\,\degC\ is sufficient to strip virtually all oxygen species from the diamond surface in vacuum. This aligns well with the saturation in termination quality observed in the forming gas annealed samples at around 850\,\degC, as we expect a reasonable deviation of the sample temperature from the thermocouple reading of the stage temperature in the UHV environment due to imperfect thermal contact between the sample and stage. There is therefore no strong evidence from these measurements to suggest that molecular hydrogen is required to catalyse oxygen removal.
\subsection{Density-Functional and Transition-State Theoretic Modelling}
\noindent To further probe the role of molecular hydrogen in the FG annealing process, we used a combination of density-functional and transition-state theory (DFT and TST) to model the dependence of hydrogen gas passivation of the reconstructed diamond (100) surface on the annealing temperature and molecular hydrogen thermal dissociation rates (Supporting Information). For DFT calculations, we used the cells shown in Figure \ref{modelling} a)-c), consisting of the first three layers of the reconstructed (100) surface restricted to three-by-three atoms when viewed in the (100) plane. We consider the case where the surface atoms are fully passivated by a hydrogen dimer (panel a), partially passivated (panel b), and in the reconstructed state with no hydrogen passivation (panel c). We denote these configurations as $\theta_0$, $\theta_1$, and $\theta_2$ respectively. For reactions describing thermal desorption of hydrogen, these being $\theta_0 \rightarrow H^{*}+\theta_1$ and $\theta_1 \rightarrow H^{*}+\theta_2$, we estimated activation barriers of 4.40\,eV and 3.46\,eV respectively. The activation barriers suggest that these reactions become significant contributors to the net surface state at temperatures higher than around 1200\,\degC. In contrast, the activation barrier for the reaction of atomic hydrogen with carbon dangling bonds was calculated to be 0\,eV. Provided atomic hydrogen is present then, such as in plasma-based termination processes, we expect hydrogen passivation of the reconstructed (100) surface to proceed rapidly across all experimentally accessible temperatures. The third key family of reactions studied were the cracking of molecular hydrogen on the diamond surface to form a free radical and a passivated surface site $\theta_2+H_2 \rightarrow \theta_1+H^{*}$ and $\theta_1+H_2 \rightarrow \theta_0+H^{*}$. We determined activation energies of 1.11\,eV and 0.30\,eV respectively for these reactions. Combining the calculated activation barriers with transition-state theory (Supporting Information) we then calculated steady-state proportions of the three surface configurations $\theta_i$ as a function of the atomic hydrogen fraction $a_{H*}$ across a temperature range from 530\,\degC\ to 1730\,\degC. The results for configurations $\theta_0$, $\theta_1$, and $\theta_2$ are respectively depicted in Figure \ref{modelling} d)-f), where the maximum temperature of 1100\,\degC\ reached in our experiment is denoted by a vertical dashed line. $a_{H*} = 0$ represents the situation where all gas-phase hydrogen is molecular, which we expect to be similar to our annealing process, while higher values of $a_{H*}$ more closely represent the conditions in hydrogen plasmas. Figure \ref{modelling} d)-f) shows that near total passivisation of dangling surface bonds should occur regardless of the gas-phase presence of atomic hydrogen species for temperatures between 530\,\degC\ and 1100\,\degC, implying that dangling surface bonds readily catalyze the cracking of molecular hydrogen. As sample temperature increases beyond 1100\,\degC, thermal hydrogen desorption is predicted to play an increasing role in determining the make-up of the surface and will lead to increased proportions of dangling bond sites. These sites could potentially translate to increased surface damage (etching) during annealing, which could be investigated in future using higher temperature annealing than was available here.\\
\newline
\noindent The temperature dependence of the hydrogen surface fraction observed in our annealing experiment is inconsistent with the above model, which predicts effectively complete hydrogenation at all temperatures investigated. However, the correlation between the observed saturation in hydrogenation quality of FG annealed samples at around 900\,\degC\ with the near-complete disappearance of the O1s peak suggests the degree of hydrogenation is limited by thermal desorption of oxygen. The similar saturation temperatures is commensurate with the activation barrier for oxygen species abstraction not being significantly modified by the presence of a hydrogen atmosphere.
\subsection{Impact on Near-Surface Nitrogen-Vacancies}
\begin{figure*}[h]
\centering
\includegraphics[width=\textwidth]{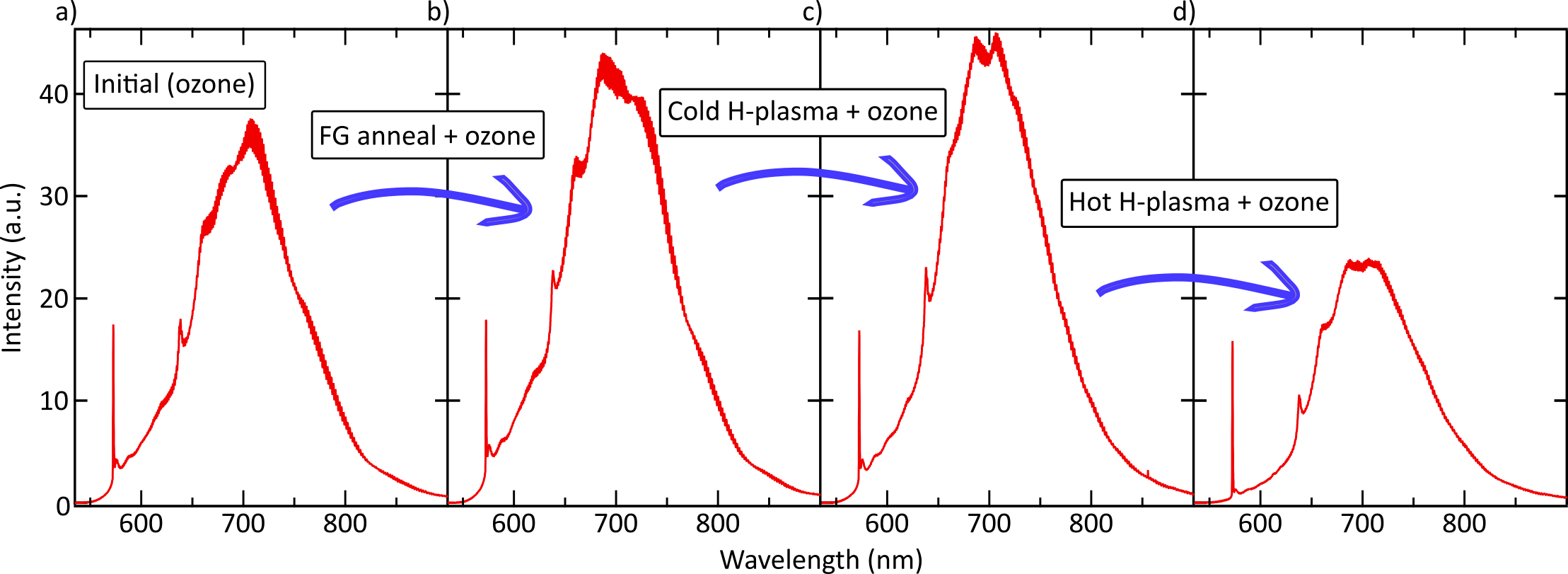}
\caption{Photoluminescence spectroscopy of near-surface nitrogen-vacancy ensemble under different surface hydrogenation treatments. a) Emission spectrum following ozone exposure prior to hydrogenation processes. b) Emission spectrum following hydrogenation by forming gas annealing and oxygen termination by ozone exposure. c) Spectrum following 55-minute unheated-stage hydrogen plasma termination followed by ozone exposure. d) Spectrum following hydrogen plasma exposure with 800\degC stage heating for 15 minutes followed by ozone treatment. All spectra were taken under identical conditions and are normalized to the first-order diamond Raman emission peak at $\approx$573\,nm, where Raman peak heights were obtained through background-subtracted Gaussian fits. Spectra from intermediate steps, showing photoluminescence emissions from the sample immediately following hydrogenation steps, are included as Supporting Information.}
\label{NVPL}
\end{figure*}

\noindent To test the suitability of both FG annealing and cold plasma exposure for hydrogenating NV containing diamond, we repeatedly hydrogen and oxygen terminated an electronic-grade sample containing an ultra-shallow NV ensemble (Methods). Figure \ref{NVPL} a) shows the baseline PL spectrum of the oxygen terminated diamond sample. We treated the sample by forming gas annealing at 950\,\degC\ for 1 hour followed by sequential 5 minute and 50 minute cold hydrogen plasma exposures and a 10 minute hot hydrogen plasma exposure. The sample was re-oxygen terminated by ozone in between each of these hydrogen termination processes, and a PL spectrum was measured following each surface treatment (both hydrogenation and oxygenation) to assess any permanent effects on the NV fluorescence. Oxygen terminated PL spectra following FG annealing, the combined 55 minutes of cold hydrogen plasma, and the 10 minutes of hot hydrogen plasma are shown in Figures \ref{NVPL} b)-d) respectively. Spectra were also measured from the sample immediately following each hydrogenation step, which exhibit the expected loss of fluorescence and conversion of NV\textsuperscript{-} to NV\textsuperscript{0} (Supporting Information) \cite{hauf_chemical_2011}. The spectra show no clear loss of NV PL after FG annealing or cold plasma termination, even with repeated treatments, highlighting the suitability of these two techniques for preserving the properties of near-surface NV centers. In stark contrast, a 10 minute heated sample plasma treatment causes an irreversible loss of almost 50\% of NV fluorescence, which is consistent with prior reports of hydrogen plasma damage to NV centers \cite{stacey_depletion_2012,findler_indirect_2020}. Finally, we note that the changes in the structure of the NV phonon side-band visible in Figure \ref{NVPL} following each surface treatment may be a result of changes to the surface vibrational density-of-states \cite{lafosse_density--states_2006} due to uncontrolled variations in the makeup of the surface oxygen termination or defect density.
\section{Conclusions}
\begin{figure*}[h]
\centering
\includegraphics[width=\textwidth]{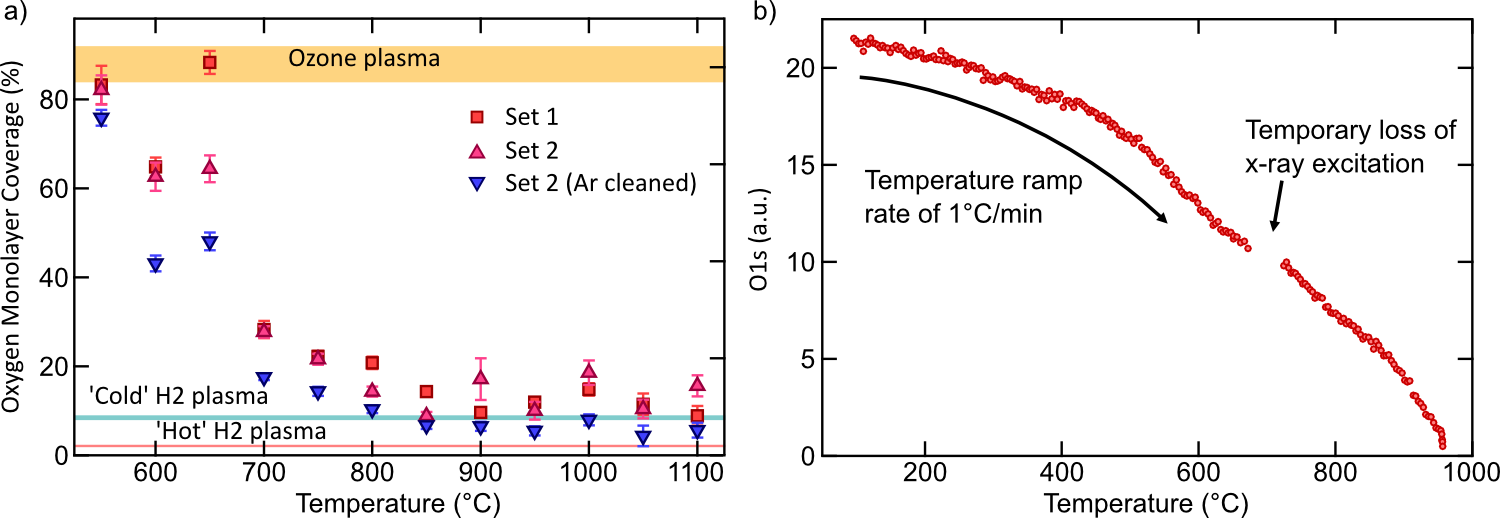}
\caption{X-ray photoelectron spectroscopy measurements of annealed diamond samples. a) Surface oxygen content, calculated using the total areas of the O1s to C1s emission peaks, as a function of hydrogen-nitrogen forming gas annealing temperature. Two sets of samples, denoted `Set 1' and `Set 2', were subjected to the same annealing processes simultaneously and are denoted by square and upward-pointing triangular markers respectively. Downward-pointing triangular markers denote measurements of Set 2 following cleaning by \textit{in situ} argon cluster ion bombardment. Shaded areas indicate the upper and lower bounds of the surface oxygen content measured on pairs of ozone treated (orange, top), `cold' hydrogen plasma treated (blue, lower), and `hot' hydrogen plasma treated (red, bottom) reference samples, all of which were cleaned by argon cluster ion bombardment prior to measurement (Methods). b) X-ray photoelectron spectroscopy measurement of the O1s peak area taken \textit{in situ} as an oxygen-terminated diamond sample underwent heating at a rate of 1\,\degC/minute in an ultra-high vacuum environment. The lack of data points around 700\,\degC\ is due to a transient loss of cooling water which caused a loss of x-ray excitation.}
\label{XPSdata}
\end{figure*}

\begin{figure*}[h]
\centering
\includegraphics[width=\textwidth]{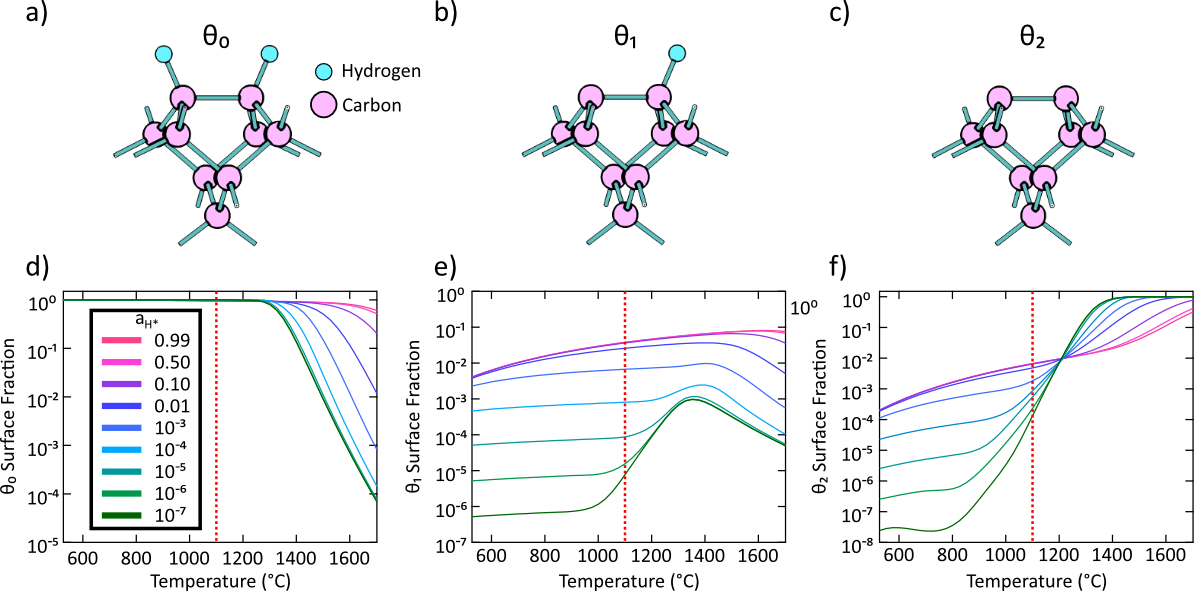}
\caption{Modelling of reconstructed diamond surface passivation in a mixed molecular and radical gaseous hydrogen atmosphere. a), b), and c) illustrate the cells $\theta_0$, $\theta_1$, and $\theta_2$ considered for modelling the fully-terminated, partially-passivated, and bare reconstructed diamond surfaces respectively. d), e), and f) show the relative fraction of the $\theta_0$, $\theta_1$, and $\theta_2$ states respectively, calculated as a function of annealing temperature for a range of atomic hydrogen fractions $a_{H*}$. For these calculations, the sample is assumed to be annealed in hydrogen gas with a partial pressure of 10\,torr. The vertical dashed lines indicate 1100\,\degC, the maximum annealing temperature experimentally reached in this work.}
\label{modelling}
\end{figure*}

\noindent Although all three approaches studied here produce conductive hydrogen terminations capable of strongly suppressing near-surface NV fluorescence by conversion to the NV\textsuperscript{0} and NV\textsuperscript{+} charge states (Supporting Information), Figure~\ref{XPSdata} a) shows that established heated sample plasma termination processes produce slightly better hydrogen coverage. One possible explanation is that the initial oxygen termination of the sample is not completely removed during FG annealing or cold plasma exposure. However, this is unlikely to be the limitation for FG annealing as the vacuum experiment presented in Figure~\ref{XPSdata} b) indicates that complete oxygen removal only requires temperatures above 950\,\degC, while annealing here was performed up to 1100\,\degC. Another possibility is that molecular hydrogen may not sufficiently fill bare diamond surface sites, leaving exposed dangling bonds until the sample is exposed to atmosphere, although we also view this as unlikely given that the modeling shown in Figure \ref{modelling} d)-f) indicates molecular hydrogen should passivate dangling bond sites immediately following oxygen abstraction.\\
\newline
\noindent For FG annealing terminations, a more likely explanation for the relatively lower hydrogen coverage is the replacement of terminating hydrogen by residual oxygen contamination in the FG flow during the short period after the sample is removed from the hot zone of the furnace and before its temperature drops below 230\,\degC, approximately the temperature at which hydrogen terminations become stable to oxygen exposure on (100) diamond surfaces \cite{maier_origin_2000}. In support of this, we note that strict control over the residual oxygen content of the FG gas flow (below 10\,ppm as measured by our RGA) was vital to consistently obtaining hydrogen terminations, and may explain why FG annealing has thus far not been widely adopted as a method for hydrogen termination of single crystal diamond. This explanation does not apply to cold plasma terminations, which may be limited by the ability of hydrogen plasma to remove certain oxygen species from unheated diamond or by the presence of trace oxygen in the plasma. Another key result is that cold plasma terminations do not result in NV passivisation, a phenomenon currently thought to result from atomic hydrogen or proton diffusion into the diamond which leads to the formation of non-fluorescent nitrogen-vacancy-hydrogen centers \cite{glover_hydrogen_2003, stacey_depletion_2012}.\\
\newline
\noindent We have examined two successful methods for hydrogenation of (100) single-crystal diamond surfaces which are non-destructive to near-surface ensembles of NV centers: Annealing of diamond above 900\,\degC\ in a high purity hydrogen-nitrogen forming gas, and hydrogen plasma exposures wherein the sample is kept below 120\,\degC. In both these cases, meticulous control of oxygen contaminants is necessary, with oxygen concentrations above 10 ppm resulting in poor hydrogenation and non-conductive surfaces. XPS reveals that both methods explored result in slightly lower hydrogen coverage compared to the standard approach using a hydrogen plasma and a heated sample stage. Modelling suggests that this is not due to an insufficient hydrogenation rate of dangling bonds on the diamond (100) surface and is more likely a result of inefficient removal of strongly-bound oxygen species such as ether groups \cite{dontschuk_x-ray_2023}.\\
\newline
\noindent The approaches studied here provide a repeatable hydrogen termination pathway suitable for developing, for instance, molecular sensing schemes that require near-surface NV centers in the vicinity of complex surface molecular functionalizations \cite{rodgers_diamond_2024}, for improving voltage imaging sensors which exploit charge state conversion in near-surface NV centers \cite{mccloskey_diamond_2022}, and for tailoring the charge states of other shallow color-centers such as SiV defects \cite{zhang_neutral_2023}.\\
\newline
\noindent \textbf{Acknowledgments} \par
\noindent We acknowledge support from the Australian Research Council (ARC) through grants DP200103712, CE170100012, and FL130100119. D.J.M. acknowledges support from a University of Melbourne proof-of-concept grant and the ARC Centre of Excellence in Quantum Biotechnology through project number CE230100021. L.V.H.R. acknowledges support by the United States National Defense Science and Engineering Graduate Fellowship. \textit{In situ} annealing and spectroscopy studies at Princeton were primarily supported by the U.S. Department of Energy, Office of Science, Office of Basic Energy Sciences, under Award No. DESC0018978, and the instrumentation development was supported by the NSF CAREER program Grant No. DMR1752047. This material is based upon work supported by U.S. Department of Energy, Office of Science, Office of Fusion Energy Sciences and Office of Basic Energy Sciences under Award No. LAB 21-2491. This work was performed in part at the Melbourne center for Nanofabrication (MCN) in the Victorian Node of the Australian National Fabrication Facility (ANFF) and also in part at RMIT University’s Microscopy \& Microanalysis Facility, a linked laboratory of Microscopy Australia enabled by the National Collaborative Research Infrastructure Strategy (NCRIS).
\bibliography{forminggaspaper_refs.bib}
\bibliographystyle{h-physrev}
\newpage
\FloatBarrier
\FloatBarrier
\newpage


\section*{Supplementary material}

# Supporting Information for

## Methods for Color Center Preserving Hydrogen-Termination of Diamond

D. J. McCloskey[1], D. Roberts[2], L.V.H. Rodgers[3], Y. Barsukov[4], I. D. Kaganovich[4], D. A. Simpson[1], N. P. de Leon[3], A. Stacey[2,5*] and N. Dontschuk[1]

[1]School of Physics, University of Melbourne; Parkville, VIC 3010, Australia
[2]School of Science, RMIT University; Melbourne, VIC 3001, Australia
[3]Department of Electrical and Computer Engineering, Princeton University, Princeton, New Jersey 08544, USA
[4]Princeton Plasma Physics Laboratory, Princeton University, Princeton, New Jersey 08543, USA
[5]Princeton Plasma Physics Laboratory, 100 Stellarator Road, Princeton, New Jersey 08543, USA

## Contents

## Modelling of Bare Diamond Surface Hydrogenation

A quantum chemistry model was used to study the hydrogenation/dehydrogenation of (100) reconstructed diamond surfaces. Reaction paths were calculated using WB97X-D DFT (density functional theory) functional and basis sets of 6-31+G(d) C atoms and 6-311+G(d,p) for H atoms. The rate constants and probabilities were calculated under transition state theory and the Eyring equation [1]; the details can be found in references [2,3]. According to equations (1) and (2), the rate constants $r(T)$ and the unitless probability of the reactions $\gamma(T)$ can be calculated via the partition function $Z$ and activation energy $E_a$:

$$(T) = A \left(\frac{T}{T_0}\right)^n \exp\left(-\frac{E_a}{RT}\right) = \frac{k_B T}{h} \frac{Z_{vib}^{TS}}{Z_{vib}^{surf} Z_{tot}^{gas}} \tag{1}$$

$$\gamma(T) = \gamma_0 \left(\frac{T}{T_0}\right)^n \exp\left(-\frac{E_a}{RT}\right) = \frac{\sqrt{2\pi m k_B T}}{h} \frac{Z_{vib}^{TS}}{Z_{vib}^{surf} Z_{tot}^{gas}} \rho_s \tag{2}$$

where $Z_{vib}^{TS}$, $Z_{vib}^{surf}$, and $Z_{tot}^{gas}$ are the vibration partition functions of the transition state (TS), surface, and the total partition function of the gas reactant respectively; $k_B$ is the Boltzmann constant, $T$ is the substrate temperature, $h$ is the Planck constant, $m$ is the mass of the gas phase reactant, and $\rho_s$ is the areal density of surface sites. All calculations were performed using the Gaussian 16 software [4].

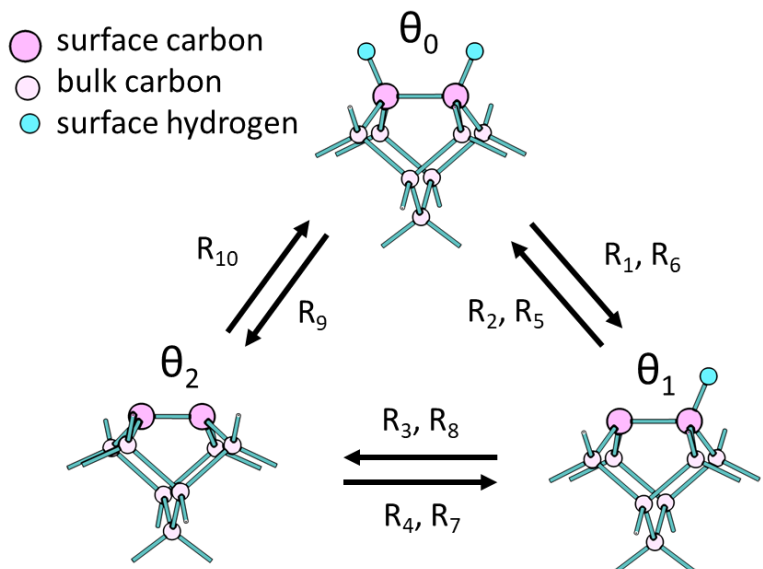

**Figure S1. Reaction diagram for (100) surface hydrogenation model.** The chemical reaction network shows the considered transformations between θ₀, θ₁, and θ₂ surface configurations, where θ₀ denotes the fully hydrogenated carbon dimer, θ₁ the half-hydrogenated dimer, and θ₂ the dehydrogenated dimer. The indicated reactions $R_i$ along with their probabilities and rate constants are given in Table S1.

The $C_9H_{14}$ cluster mimics the fully hydrogenated (100) reconstructed diamond surface consisting of carbon dimers. Three surface configurations, the fully hydrogenated carbon dimer (θ₀), the half-hydrogenated dimer (θ₁), and the dehydrogenated dimer (θ₂), were considered. The chemical reaction network describing transformations between θ₀, θ₁, and θ₂ surface configurations are shown in Figure S1. The reactions studied and their calculated probabilities and activations energies are presented in Table S1.

**Table S1. Calculated reaction probabilities and activation energies.** The list of reactions is considered under the kinetic model describing the transformation of $\theta_0$, $\theta_1$, and $\theta_2$ surface sites. The unitless $\gamma_0$ or A (in Hz), n, and $E_a$ are used in equations (1) and (2).

| $R_i$ | Reaction | $\gamma_0$ or A (Hz) | n | $E_a$ (eV) |
|---|---|---|---|---|
| 1 | $H + \theta_0 \rightarrow H_2 + \theta_1$ | 0.2987 | 0.50 | 0.33 |
| 2 | $H_2 + \theta_1 \rightarrow H + \theta_0$ | 0.1729 | 0.00 | 0.30 |
| 3 | $H + \theta_1 \rightarrow H_2 + \theta_2$ | 0.2939 | 0.55 | 0.16 |
| 4 | $H_2 + \theta_2 \rightarrow H + \theta_1$ | 0.1817 | 0.00 | 1.11 |
| 5 | $H + \theta_1 \rightarrow \theta_0$ | 1 | 0 | 0.00 |
| 6 | $\theta_0 \rightarrow H + \theta_1$ | $8.59 \times 10^{13}$ Hz | 2.30 | 4.40 |
| 7 | $H + \theta_2 \rightarrow \theta_1$ | 1 | 0 | 0 |
| 8 | $\theta_1 \rightarrow H + \theta_2$ | $8.59 \times 10^{13}$ Hz | 2.30 | 3.46 |
| 9 | $\theta_0 \rightarrow H_2 + \theta_2$ | $6.76 \times 10^{12}$ Hz | 1.40 | 5.67 |
| 10 | $H_2 + \theta_2 \rightarrow \theta_0$ | 0.1014 | -1.00 | 2.21 |

The set of equations describing the transformation between $\theta_0$, $\theta_1$, and $\theta_2$ surface configurations according to the list of the reactions in Table S1 is shown below:

$$\frac{d\theta_0}{dt} p_s = -\gamma_1 J_H \theta_0 + \gamma_2 J_{H2} \theta_1 + \gamma_5 J_H \theta_1 - k_6 p_s \theta_0 - k_9 p_s \theta_0 + \gamma_{10} J_{H2} \theta_2$$

$$\frac{d\theta_1}{dt} p_s = \gamma_1 J_H \theta_0 - \gamma_2 J_{H2} \theta_1 - \gamma_3 J_H \theta_1 + \gamma_4 J_{H2} \theta_2 + \gamma_5 J_H \theta_1 - k_6 p_s \theta_0 + \gamma_7 J_H \theta_2 - k_8 p_s \theta_1$$

$$\theta_0 + \theta_1 + \theta_2 = 1, \tag{3}$$

where $J_H$ and $J_{H2}$ in the set of equations (3) are thermal fluxes of H atoms and $H_2$ molecules:

$$J_{H2} = \frac{1}{4}\sqrt{\frac{8k_BT}{m\pi}} * \frac{P(H_2)}{RT}$$

$$J_H = \frac{1}{4}\sqrt{\frac{8k_BT}{m\pi}} * \frac{P(H)}{RT}. \tag{4}$$

The set of equations (3) were solved under steady-state conditions to yield the fractions of surface configurations shown in the main text.

# Full Series of Hydrogenation Treatments of Near-Surface NV Ensemble

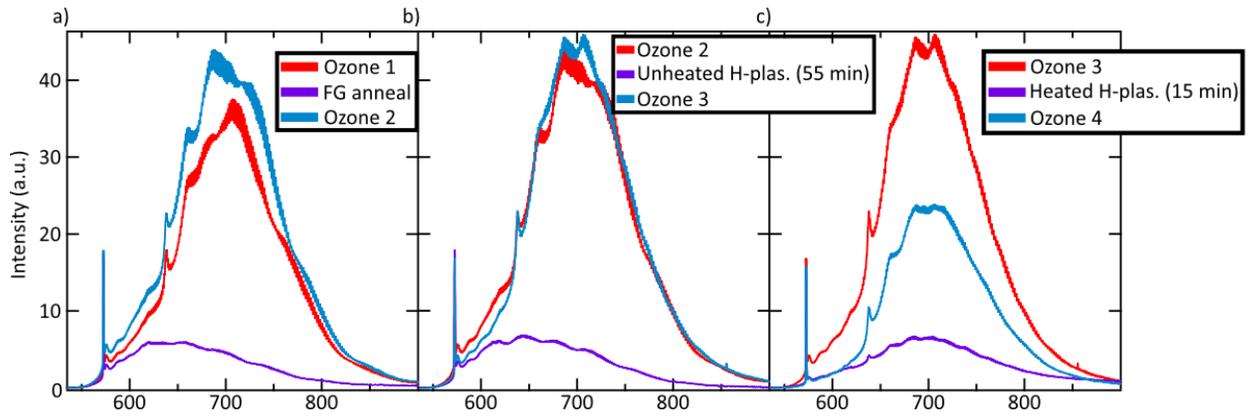

**Figure S2. Photoluminescence of shallow nitrogen-vacancy ensemble subject to successive hydrogen- and-oxygen-termination treatments.** a) Series of photoluminescence spectra following forming gas annealing at 950°C for 1 hour. b) Series of spectra following hydrogen plasma with no sample stage heating for 55 minutes. c) Series of spectra following hydrogen plasma treatment with 800°C sample stage heating for 15 minutes.

We performed a series of alternating oxygen-and-hydrogen termination treatments on a diamond sample containing a dense, near-surface ensemble of NV centers (see main text for sample details) in order to assess the destructive potential of these treatments on the ensemble photoluminescence (PL) intensity, which serves as a proxy for the NV density.

Beginning with Figure S2a, we first probed the effects of forming gas annealing. Starting with an ozone treatment of the sample to establish a consistent oxygen termination, an initial PL spectrum (red) revealed bright NV fluorescence heavily weighted towards the negative (NV$^-$) charge state. The sample was annealed in forming gas at 900°C for 1 hour, after which the spectrum (purple) showed a dramatic reduction in PL intensity and an apparent total conversion to the neutral (NV$^0$) charge state, evidenced by the loss of the NV- zero-phonon line (ZPL) at 637nm and the shift of the phonon sideband towards a peak value around 650nm, as would be expected for hydrogen-termination of very shallow NV ensembles [5]. The sample was again treated with ozone, where the resulting spectrum (blue) displayed a slight increase in PL intensity over the initial state of the sample but a similar charge state ratio consistent with FG annealing at 900°C being non-destructive to near-surface NV centers. The slight variation in intensity relative to the diamond Raman line may be attributed to slight variation in the focus of the confocal spectrometer.

Next, we examined hydrogen plasma treatments without stage heating (Figure S2b). Beginning with the sample immediately following the set of FG annealing measurements (red), we exposed the sample to hydrogen plasma for 55 minutes (see main text for detailed recipe). The resulting PL (purple) showed the formation of a good hydrogen termination as evidenced by total conversion to NV$^0$ and significantly lower PL overall. The return of the PL to baseline level following ozone treatment (blue) despite 55 minutes of hydrogen plasma exposure supports the conclusion that plasma termination without a heated sample stage preserves near-surface NV centers.

Finally, we exposed the sample to hydrogen plasma for 15 minutes while heating the sample stage to a set temperature of 800°C (full recipe detailed in main text) and followed the resulting changes to PL emission (Figure S2c). PL from the sample following plasma treatment (purple) showed a large reduction in absolute intensity when compared to ozone treatment (red). Unlike the FG annealed or cold-plasma

treated samples, the NV⁻ ZPL at 637nm persisted following heated sample hydrogen plasma treatment, together with a more pronounced phonon sideband peaking at 700nm. Following another ozone treatment, PL emissions from the NV ensemble only returned to around half of their original intensities (blue). We additionally confirmed that further ozone treatments could not restore the PL further.


\end{document}